# Quantum optics and cavity QED with quantum dots in photonic crystals

Jelena Vučković, Stanford University

This chapter will primarily focus on the studies of quantum optics with semiconductor, epitaxially grown quantum dots (QDs) embedded in photonic crystal (PC) cavities. Therefore, we will start by giving brief introductions into photonic crystals and quantum dots, although the reader is advised to refer to other references (John D. Joannopoulos 2008) (Michler 2009) for more detailed studies of these topics. We will then proceed with the introduction to cavity quantum electrodynamics (QED) effects (Kimble, Cavity Quantum Electrodynamics 1994) (Haroche 2013), with a particular emphasis on the demonstration of these effects on the quantum dot-photonic crystal platform. Finally, we will focus on the applications of such cavity QED effects.

1. **Photonic crystals and microcavities**

   **1.1. Photonic crystals**

Photonic crystals are media with periodic modulation of dielectric constant in up to three dimensions (John D. Joannopoulos 2008) (Yablonovitch 1987) (John 1987). In the literature, the name "photonic crystals" usually refers to the structures with dielectric constant periodic in two and three dimensions, while one-dimensional periodic media are referred to as the distributed Bragg reflectors (DBRs).

Let us consider a non-magnetic periodic medium (the permeability is equal to $\mu_0$), described with a periodic dielectric constant $\varepsilon(\vec{r}) = \varepsilon(\vec{r} + \vec{a})$, where $\vec{a}$ is an arbitrary lattice vector. The allowed electromagnetic modes in such a medium can be obtained as the solutions of the wave equation:
$$\vec{\nabla} \times \vec{\nabla} \times \vec{E} = \omega^2 \varepsilon(\vec{r}) \mu_0 \vec{E},$$
where the dielectric constant is periodic $\varepsilon(\vec{r}) = \varepsilon(\vec{r} + \vec{a})$, as described above.

Similarly to the wavefunction of an electron inside a periodic potential imposed by the crystal lattice, it can be shown that electromagnetic field inside such a periodic dielectric medium can be expanded in terms of the electromagnetic eigenmodes which satisfy the Bloch theorem, and are therefore referred to as the Bloch modes (Bloch states) (Yeh 2002). The electric field of the three dimensional Bloch modes can be described as:
$$\vec{E}_{\vec{k}}(\vec{r}) = e^{i\vec{k}\cdot\vec{r}} \vec{u}_{\vec{k}}(\vec{r})$$
$$\vec{u}_{\vec{k}}(\vec{r} + \vec{a}) = \vec{u}_{\vec{k}}(\vec{r})$$

where $\vec{k}$ is the wave-vector, and $\vec{u}_{\vec{k}}(\vec{r})$ is the periodic part of the Bloch state.

Therefore, given the frequency ω, we can solve the wave equation in such a periodic medium to find the wavevector $\vec{k}$ and the allowed Bloch modes of the system at that frequency. The relation $\omega(\vec{k})$ is called the *photonic band diagram* (or *dispersion relation*).

If there is a whole band of frequencies between $\omega_{min}$ and $\omega_{max}$ in which the wave equation inside such a periodic crystal has no solutions for any real $\vec{k}$, such a band of frequencies is called the *forbidden (stop) band*, or the *photonic band gap*. Practically, the Bloch modes in this range of frequencies do not exist, implying that the electromagnetic waves with frequencies within the photonic band gap cannot propagate through the photonic crystal. In other words, the photonic crystal behaves as a mirror for electromagnetic waves with frequencies inside the photonic band gap. By analogy with the study of one dimensional (1D) periodic media, such reflection could also be viewed as the distributed Bragg reflection of the electromagnetic wave from the interfaces between regions with different dielectric constants. As in the calculation of band diagrams in solid-state physics, to determine the size of the photonic band gap, it is sufficient to only solve for the photonic band diagram inside the *irreducible first Brillouin zone* (this is the result of the band diagram's periodicity and symmetry in k-space).

In photonic crystals which have the dielectric constant periodic in less than three dimensions, a photonic band gap can occur, but only for the waves propagating in a certain direction or set of directions in which the periodicity of the dielectric constant occurs. Three-dimensional (3D) photonic crystals, however, can lead to a complete photonic band gap, meaning that in a certain frequency region, the wave propagation is prohibited through the crystal in 1) any direction in space and 2) for any polarization.

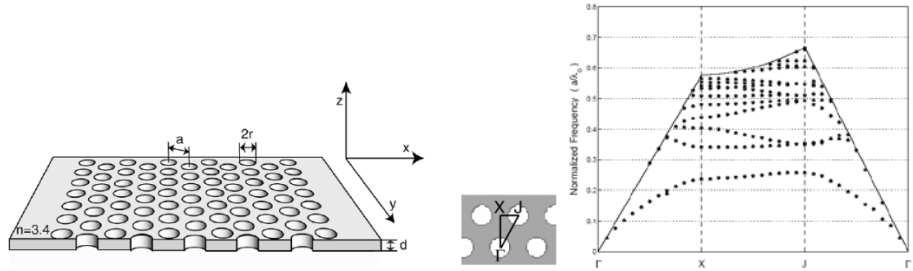

**Figure 1. Left:** Optically thin membrane patterned with a hexagonal array of air holes. The inset illustrates the orientation of the principal directions in the reciprocal space. **Right:** Band diagram for TE-like modes of a thin slab surrounded by air on both sides and patterned with a hexagonal array of air holes. Parameters of photonic crystal are *n*=3.4, *d/a*=0.65 and *r/a*=0.3.

Although 3D photonic crystals offer the opportunity for light manipulation in all three dimensions in space, they are very difficult to fabricate. For this reason, most of the research efforts in the past have been focused on planar photonic crystals, i.e., one- and two-dimensional (2D) photonic crystals of finite depth, which can be made by standard microfabrication methods. The light confinement in planar photonic crystals results from the combined action of the distributed Bragg reflection in the photonic crystal and the total internal reflection (TIR) in the remaining dimensions. The imperfect confinement by TIR produces some unwanted radiation loss, which is usually a limiting factor in performance of these structures; still, most of the functionality of 3D photonic crystals can be achieved by careful design.

There are five parameters of planar photonic crystals that one can control, as illustrated in Figure 1: the refractive index of material (*n*), the type of photonic crystal lattice (hexagonal, square...), the thickness of the slab (*d*), the lattice periodicity (*a*), and the hole radius (*r*). The wavelength of light in air is denoted as $\lambda_0$. Refractive index is typically in the range between 3.4 and 3.6, for photonic crystals made out of semiconductors such as Si, GaAs, and InP and operating at near-infrared wavelengths. In vertically symmetric planar photonic crystals, modes can be classified either as vertically even, i.e., transverse electric (TE)-like, or vertically odd, i.e., transverse magnetic (TM)-like. At the center of the photonic crystal slab (center in the z-direction, labeled here as z=0), TE-like modes look exactly like the TE-modes of the 2D photonic crystals (infinite in z-direction): the only non-zero field components they have are $H_z$, $E_x$, and $E_y$. Similarly, the TM-like modes are the same as the TM modes at the center of the slab, and have only $E_z$, $H_x$, and $H_y$ nonzero field components. Away from the center of the slab, both TE-like and TM-like modes contain all E and H field components, i.e., are not purely TE or TM; this is a result of the mode confinement in the vertical direction. It should be also pointed out that the TE-like and TM-like modes are sometimes referred to as even and odd, respectively, based on whether they exhibit even or odd mirror symmetry in the vertical direction, at the center of the slab. Even mirror symmetry means that $H_z$, $E_x$, and $E_y$ are even at z=0, while the remaining three field components are odd, while the odd mirror symmetry means that $E_z$, $H_x$, and $H_y$ are even at z=0, and the remaining three field components are odd. This is a result of the application of an even or odd mirror plane at z=0, for which it holds that $M_z \vec{E} = \sigma_z \vec{E}$, with $\sigma_z=1$ or $\sigma_z=-1$, respectively (the mirror operator $M_z$ is defined as $M_z \vec{E} = M_z(E_x, E_y, E_z) = (E_x, E_y, -E_z)$).

The photonic band diagram (dispersion diagram) of TE-like modes of the planar photonic crystal with a hexagonal lattice of air holes and with parameters *n*=3.4, *d/a*=0.65 and *r/a*=0.3 is shown in Figure 1. On the horizontal axis, we plot the values of the wavevector in plane, and on the vertical axis we plot the normalized frequency of light in units of $a/\lambda_0$. These calculations have been done by using the first-principles 3D finite-difference time-domain (FDTD) method, which is based on the discretization of the Maxwell's equations in space and time. This band diagram is plotted only along the high-symmetry directions of the irreducible first Brillouin zone (ΓX, XJ and ΓJ). These directions are labeled along the x-axis of the band diagram shown in the figure.

In the band diagrams of 2D periodic structures like the one in Fig. 1 we plot only the in-plane value of the wave-vector $k_{//}=(k_x,k_y)$ on the horizontal axis, where $|k_{//}|^2= k_x^2+k_y^2$. However, the total wave-vector also has a $k_z$ component and we can write $k_x^2+k_y^2+k_z^2 =k^2=(\omega/c)^2$. A wave that is free to propagate in the z-direction (*leaky mode*) would have a real $k_z$, i.e., its z-dependence can be expressed as a plane wave $e^{ik_z z}$, where $k_z$ is real. On the other hand, a wave that is confined by TIR in the z-direction (*guided mode*) would decay exponentially in that direction (be evanescent in the z-direction) and therefore have an imaginary $k_z$. Therefore, if we plot the *light line* $\omega_{ll}=c|k_{//}|$ (corresponding to a wave without any $k_z$) on the band diagram, then for an evanescent mode we would have $k_x^2+k_y^2+k_z^2 = (\omega/c)^2<(\omega_{ll}/c)^2$, and for a leaky mode we would have $k_x^2+k_y^2+k_z^2 = (\omega/c)^2>(\omega_{ll}/c)^2$. Therefore, modes below the light line are confined by TIR, can be guided in the slab and are called *guided modes* (they are evanescent in the *z*-direction and are confined by total internal reflection), whereas modes above the light line (in the gray region; individual modes not plotted in the figure) are called *leaky modes*.

In Fig. 1, one can also observe the frequency region where guided modes do not exist, for any value of the in-plane wave-vector; this is the *photonic bandgap*. Guided modes are organized into modes above the bandgap (*air band*), and modes below the bandgap (*dielectric band*); the names "dielectric" and "air" band are based on where the electric field energy of a mode is mostly concentrated. For example, the dielectric band modes mostly concentrate their electric field energy in the high refractive index region (semiconductor), while the air band modes mostly concentrate their electric field energy in air region (holes) (John D. Joannopoulos 2008). The air band is sometimes also referred to as the *conduction band*, and the dielectric band is sometimes referred to as the *valence band*, by analogy with energy band diagrams of semiconductors.

### 1.2. Planar photonic crystal microcavities

As discussed previously, two-dimensional photonic crystals of finite depth can exhibit a photonic band gap for electromagnetic waves propagating in the plane of the crystal. Inside the photonic band gap, the density of optical states is zero; outside the band gap, Bloch modes exist that can be classified based on their k-vector. However, by perturbing a photonic crystal lattice (i.e., by introducing lattice defects), one can permit localized modes that have frequencies within the photonic band gap. Such modes have to be evanescent inside the photonic crystal, i.e., they have to decay exponentially away from the defect. In other words, the defect behaves as an optical cavity, and the surrounding photonic crystal represents mirrors surrounding the cavity. Therefore, the defects introduce peaks into the density of optical states inside the photonic band gap. Moreover, the defects break the discrete translational symmetry of the photonic crystal, and one can no longer classify the modes based on their k-vector.

The simplest way of forming a microcavity starting from the unperturbed hexagonal photonic crystal lattice of air holes, for example, is by changing the radius of a single hole, or by changing its refractive index (Jelena Vuckovic 2002). The former method is more interesting from the perspective of fabrication. By simply increasing the radius of a single hole, an acceptor defect state is excited, i.e., pulled into the bandgap from the dielectric band. On the other hand, by decreasing the radius of an individual hole (or by tuning its refractive index between 1 and the refractive index of the slab), a donor defect state is excited and pulled into the bandgap from the air band, as shown in Figure 2. It is also important to note that in this section, we are focusing only on TE-like modes, for which a band gap exists in such planar photonic crystal.

The two most important parameters characterizing the cavity mode (in addition to its wavelength, or frequency) are its quality (Q) factor and the mode volume (V). Q factor describes the storage time of a photon (or classically, of electromagnetic field energy) inside a cavity; the energy stored inside the mode with frequency ω and quality factor Q would decay in time (t) as : $W(t) = W_0 e^{-\frac{\omega t}{Q}} = W_0 e^{-2\kappa t}$, where $\kappa = \frac{\omega}{2Q}$ is the *cavity field decay rate*. The volume of space in which the field is concentrated is described by the cavity mode volume: $V = \frac{\iiint \varepsilon(\vec{r})|\vec{E}(\vec{r})|^2 d^3\vec{r}}{\max\{\varepsilon|\vec{E}(\vec{r})|^2\}}$. Clearly, high Q increases the photon storage time, and small V increases field localization, and this is usually the range in which we prefer our system to operate in order to increase the light-matter interaction, as described later.

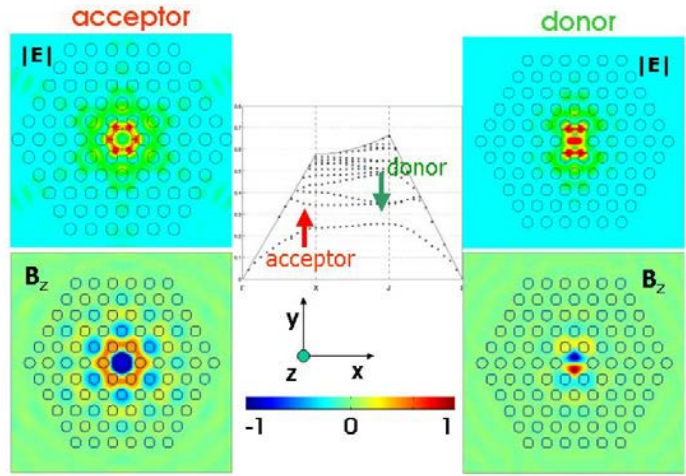

**Figure 2.** Acceptor defect state and donor defect state excited by changing the radius of a single photonic crystal hole.

Even the presented simple single defect microcavities produced by changing the radius or refractive index of a single photonic crystal hole can localize light into volumes as small as one half of a cubic wavelength in the material. However, more sophisticated designs can lead to much higher Q-factors while preserving such small mode volumes (Bong-Shik Song 2003). The highest Q-factors of photonic crystal cavities have been demonstrated in silicon at 1550nm, where material absorption is minimal. Q-factors as high as $10^6$ have been reported recently by passive, transmission measurements (Takashi Asano 2006).

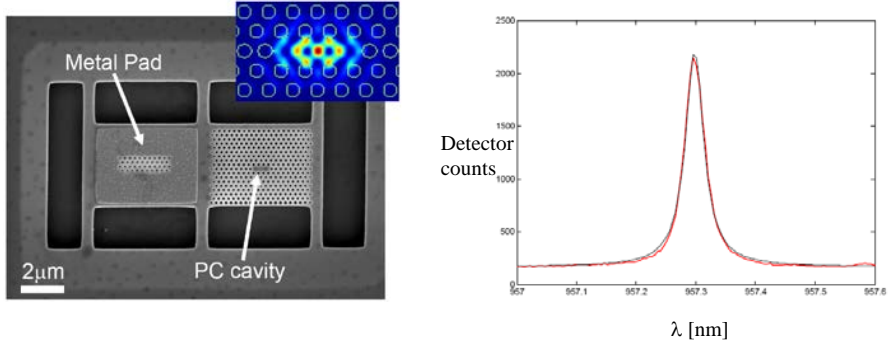

**Figure 3. Left:** L3 photonic crystal cavity fabricated in GaAs. The inset shows the simulated electric field pattern inside such a cavity. The simulated optical mode volume is V~$0.7(\lambda/n)^3$. **Right**: the measured spectrum of such a cavity with the quality factor Q=25300.

In this chapter we primarily focus on the so-called L3 photonic crystal cavity produced by removing 3 holes in hexagonal photonic crystals along a line, and perturbing some neighboring holes (Yoshihiro Akahane 2003). The scanning electron microscope (SEM) image of such a microcavity fabricated in GaAs, the electric field pattern, and the measured spectrum for one of the fabricated cavities are shown in Figure 3.

## 2. Quantum dots

### 2.1. Semiconductors

Let us consider a simple model of a solid in which the positive ions comprise a uniform array of fixed sites (periodic structure referred to as the crystal lattice) and the electrons are free to move through such a periodic lattice. This model can be used to describe semiconductors. To simplify the analysis, we will also make a so-called one electron approximation: we neglect the interaction between electrons, vibrations of the crystal, and assume that an electron moves through the crystal as if it were alone, and that it experiences a potential that has the same periodicity as the crystal lattice, which is resulting from the periodic arrangement of the cristal lattice ions. This is already a sufficient amount of information to set up the (time-independent) Schrödinger equation for such an electron:

$$-\frac{\hbar^2}{2m}\nabla^2\psi(\vec{r})+V_p(\vec{r})\psi(\vec{r})=E\psi(\vec{r}),$$

or in the form of energy eigenvalue equation:

$$H\psi(\vec{r}) = E\psi(\vec{r})$$

$$H = -\frac{\hbar^2}{2m}\nabla^2 + V_p(\vec{r})$$

The potential that an electron experiences is periodic with crystal lattice periodicity $\vec{R}_L$ is:

$$V_p(\vec{r} + \vec{R}_L) = V_p(\vec{r})$$

and m is the mass of an electron: m=9.1·10⁻³¹kg.

This is completely analogous to the wave equation for electromagnetic waves inside photonic crystals, and so the energy eigenstates for an electron are supposed to be in the form of previously introduced Bloch states:

$$\psi_{\vec{k}}(\vec{r}) = e^{i\vec{k}\cdot\vec{r}} u_{\vec{k}}(\vec{r})$$

$$u_{\vec{k}}(\vec{r} + \vec{a}) = u_{\vec{k}}(\vec{r})$$

Therefore, energy eigenstates of an electron inside a periodic potential (crystal lattice) are Bloch states, which satisfy the Bloch boundary condition given above.

Therefore, by solving this Schrödinger equation, we can obtain all possible eigenenergies of an electron inside the crystal structure described with a periodic potential $V_p(\vec{r})$. In other words, we can find its energy band diagram, i.e., the allowed combinations of eigenenergy E and the Bloch state wavevector k. In case of semiconductors, such as GaAs, the band diagram exhibits energy band gap – a region of energies that an electron cannot have inside such a material (also called a semiconductor band gap).

Like in the case of photonic crystals, in order to determine the size of the energy band gap, it is sufficient to solve the band diagram only along the edges of the irreducible 1st Brillouin zone.

**2.2. Effective mass approximation**

Let us assume that we are interested only in the region of the band diagram around the minimum or maximum at *k*=0 (e.g., bottom of conduction or top of valence band) in direct band semiconductors, with the band gap equal to V. In that case, we can make a parabolic approximation, i.e., assume that conduction and valence bands around k=0 can be approximated as parabolic:

$$E_c(k) = V + \frac{\hbar^2 k^2}{2m_{eff,c}}$$

$$E_v(k) = \frac{\hbar^2 k^2}{2m_{eff,v}}$$

$m_{eff,c}$ and $m_{eff,v}$ are effective masses of the conduction and valence bands which are positive and negative, respectively.

A localized electron state can be expressed as a superposition of the Bloch states from one band in a narrow region around *k*=0:

$$\psi(\vec{r},t) = \sum_{\vec{k}} c_{\vec{k}} u_{\vec{k}}(\vec{r}) e^{i\vec{k}\vec{r}} e^{-iE_{\vec{k}} t/\hbar}$$

For a narrow region of k in superposition, $u_{\vec{k}}(\vec{r}) \approx u_0(\vec{r})$, implying that:

$$\psi(\vec{r},t) \approx u_0(\vec{r}) \sum_{\vec{k}} c_{\vec{k}} e^{i\vec{k}\vec{r}} e^{-iE_{\vec{k}} t/\hbar} = u_0(\vec{r}) \psi_{env}(\vec{r},t)$$

where $\psi_{env}(\vec{r},t)$ is the *envelope wavefunction*.

Using the parabolic approximation to the bands, it can then be shown that the envelope function satisfies the following equation:

$$-\frac{\hbar^2}{2m_{eff}} \nabla^2 \psi_{env}(\vec{r},t) + V(\vec{r}) \psi_{env}(\vec{r},t) = i\hbar \frac{\partial \psi_{env}(\vec{r},t)}{\partial t}$$

where V(r) is the size of the band gap as a function of position. The last equation looks like the Schrödinger equation for the envelope function. This result is also-called the *effective mass approximation*. The boundary conditions used for solving this equation are a continuity of $\frac{1}{m_{eff}} \nabla \psi_{env}(\vec{r},t)$ and $\psi_{env}(\vec{r},t)$.

In the effective-mass approximation, an electron is treated as a particle with the mass equal to the effective mass $m_{eff}$ (instead of the free electron mass $m_0$=9.1·10$^{-31}$kg). We forget about the periodic part of the Bloch state, and treat only the wavefunction envelope.

### 2.3. Quantum dots

The only way to completely localize the wavefunction and obtain discrete energy levels is by localizing it in more than one dimension, i.e., by surrounding the particle by a potential barrier in all directions. This is achieved in a quantum dot. Quantum dots are made usually by sandwiching a tiny chunk of one type of semiconductor with a smaller band gap inside another semiconductor with a larger band gap, thereby producing potential barrier for electrons and holes in three dimensions.

For example, InAs quantum dots embedded inside GaAs (1μm² arrays) are shown in Figure 4. Such quantum dots are formed by self-assembly during the growth process called molecular beam epitaxy (MBE), as a result of lattice mismatch between InAs and GaAs (Michler 2009). This process of self-assembly of quantum dots is called Stransky-Krastanov growth. Since quantum dot islands are formed by self-assembly, their locations are random, and there is a distribution of sizes and shapes, leading to variation in the energies of transitions of different quantum dots on the same wafer (inhomogeneous broadening).

It is interesting to note that usually not all InAs from the thin deposited layer is turned into such islands. A very thin sheet remains under the quantum dots, which is in fact a very thin quantum well, also referred to as the wetting later. This layer somewhat affects the 3D confinement of carriers.

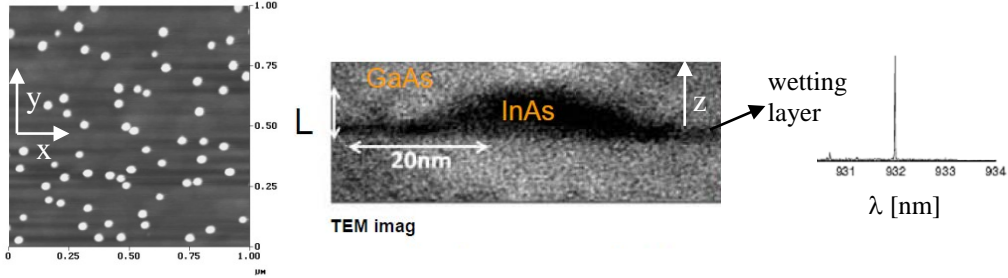

**Figure 4.** Left: Atomic force microscope (AFM) image of an uncapped InAs quantum dot array. The area shown is 1 μm x 1 μm. Middle: a transmission electron microscope (TEM) image of a cross section of an assembled InAs/GaAs quantum dot (courtesy of Prof. Jonathan Finley, TU Munich). Right: photoluminescence spectrum of a single InAs/GaAs quantum dot, showing a neutral exciton line.

Quantum dots exhibit discrete atomic like spectra (shown in Figure 4), and for this reason they are often referred to as artificial atoms. The discrete lines in the spectrum of quantum dots (as shown below) are produced by recombination of carriers (electrons and holes) occupying discrete energy levels in the conduction and valence bands.

The energy structure of the quantum dot can be found by solving the following Hamiltonian for an electron-hole pair in a quantum dot, where the subscripts e, h, and c refer to electron, hole, and Coulomb interaction, respectively:

$$H = H_e + H_h + H_c$$

$$H_e = -\frac{\hbar^2}{2}\nabla\left[\frac{1}{m_{eff,e}}\nabla\right] + V_e(\vec{r})$$

$$H_h = -\frac{\hbar^2}{2}\nabla\left[\frac{1}{m_{eff,h}}\nabla\right] + V_h(\vec{r})$$

$$H_c = -\frac{e^2}{4\pi\varepsilon}\frac{1}{|r_e - r_h|}$$

Typically, we first solve single particle problems for electron and hole ($H_e$, $H_h$) and then treat Coulomb interaction ($H_c$) as a perturbation. The potential V(r) is usually assumed to be harmonic in $x$-$y$ plane and an infinite potential well in z direction (along the QD growth axis, see Figure 4):

$$V(x,y,z) = \begin{cases} \frac{1}{2}m_{eff}\omega_0^2(x^2+y^2), & \text{for } z \leq L \\ \infty, & \text{for } z > L \end{cases}$$

This expression holds for both electrons and holes, but the appropriate parameters have to be substituted ($m_{eff,e}$, $m_{eff,h}$). $L$ is the height of the quantum dot in the z-direction (as shown in the Figure 4) and $\hbar\omega_0 = 30 - 80 meV$.

The energy levels resulting from this Hamiltonian are:

$$E = (n_x + n_y + 1)\hbar\omega_0 + \frac{1}{2m_{eff}}\left(\frac{\pi\hbar n_z}{L}\right)^2$$

and the corresponding envelope wavefunctions are:

$$\psi_{env}(x,y,z) \propto H_{n_x}(x)H_{n_y}(y)e^{-\frac{m_{eff}\omega_0}{2\hbar}(x^2+y^2)}\sin\left(\frac{\pi n_z z}{L}\right)$$

where $n_x$, $n_y$ are non-negative integers, and $n_z$ is a positive integer. In case of InAs/GaAs quantum dots, the height L is small, so that usually only $n_z=1$ is confined. $H_{nx}$ and $H_{ny}$ are Hermite polynomials. The second term in the sum for E and the last term in the product for ψ describe the potential well in z-direction, while the remaining terms describe two harmonic oscillators, in x and y directions. As described in the Section 2.2, the full wavefunction also contains a periodic part (in addition to the envelope).

It should also be noted that the expression for eigenenergies above does not include Coulomb interaction. To include the effect of Coulomb interaction ($H_c$) on energy eigenstates, we can treat $H_c$ as perturbation Hamiltonian:

$$\Delta E_0 = \langle 0|H_c|0\rangle + \sum_i \frac{|\langle 0|H_c|i\rangle|^2}{E_i - E_0}$$

where |i> represents excited eigenstates of the system without perturbation (including both electron and hole), and $E_i$ are corresponding eigenenergies. Similarly, |0> and $E_0$ correspond to ground state and the corresponding eigenergy:

$$|0\rangle \rightarrow \psi_0(r_e)\psi_0(r_h)$$
$$E_0 = E_{0e} + E_{0h}$$
$$|i\rangle \rightarrow \psi_i(r_e)\psi_i(r_h)$$
$$E_i = E_{ie} + E_{ih}$$

By solving this perturbation Hamiltonian for a typical InAs/GaAs QD, one can obtain corrections on the energy of a single electron-hole pair on the order of $\Delta E_0$= -20meV (M. Grundmann 1995). A single electron-hole pair in the quantum dot is referred to a neutral exciton (X).

In a similar way, it can be shown that when adding charges of the same sign to a quantum dot, the energy required to add each new particle increases by 10-20 meV, due to electrostatic (Coulomb) repulsion (estimated through perturbation theory). An excitonic complex consisting of two electrons and one hole is referred to as a negatively charged trion ($X^-$) etc. Following the discussion above, after recombination of various carrier complexes inside a quantum dot, different photon energies are expected at the output. This can also be seen in Figure 5, where by applying a voltage bias (electric field), the QD goes through various charge configurations, and its photoluminescence spectrum at the output changes. (Konstantinos G. Lagoudakis 2013)

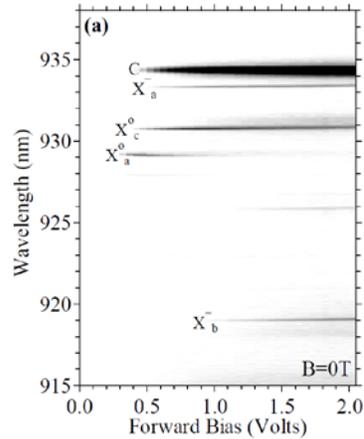

**Figure 5.** Overview of spectral lines as a function of applied bias on the nanocavity containing a few quantum dots (labeled a,b,c). The lines are labeled as: nanocavity mode C, neutral excitons $X_a^0$, $X_c^0$, and charged excitons $X_a^-$, $X_b^-$. As the QD gradually charges, the neutral exciton line disappears while the trion line appears (most visible for QD "a" beyond 1.2V, where the QD is charged with maximal probability).

## 3. Introduction to cavity quantum electrodynamics (cavity QED)

### 3.1. Quantum-mechanical treatment of the atom - electromagnetic field interaction: Jaynes-Cummings Hamiltonian

We are considering a two level emitter with excited state |ex> and ground state |g>, and the transition frequency ($\nu=E/\hbar$) between two energy levels (Fig. 6). We also assume that the system is on resonance or close to resonance with the fundamental optical cavity mode frequency ω. Under these conditions, the excitation of other cavity modes can be neglected, and the system can be modeled as a single two-level system coupled to a single cavity mode. This coupled system can be described by the Hamiltonian:

$$H=H_E+H_F+H_{int}.$$

The three terms of the Hamiltonian are the emitter Hamiltonian ($H_E$), the field Hamiltonian ($H_F$), and the emitter-field interaction Hamiltonian ($H_{int}$).

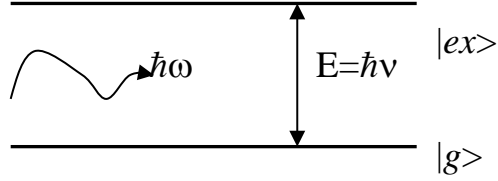

**Figure 6.** Two level system interacting with a single mode of electromagnetic field.

The emitter Hamiltonian is $H_E = \frac{\hbar \nu}{2}\sigma_z$, where $\sigma_z = |ex\rangle\langle ex| - |g\rangle\langle g|$ is the population operator. |ex> and |g> represent a complete set of energy eigenstates for the emitter, with corresponding eigenvalues $\hbar\nu/2$ and $-\hbar\nu/2$, respectively. They also represent a basis for the emitter, i.e. the closure relation holds $1 = |ex\rangle\langle ex| + |g\rangle\langle g|$. The field Hamiltonian is $H_F = \hbar\omega\left(a^\dagger a + \frac{1}{2}\right)$, where $a$ and $a^\dagger$ are photon annihilation and creation operators, respectively. The eigenstates of $H_F$ are Fock (photon number) states $|n\rangle$, with corresponding eigenvalues $\hbar\omega\left(n + \frac{1}{2}\right)$.

The interaction Hamiltonian between the emitter and the cavity field can be described semi-classically in the dipole approximation:
$$H_{int} = -\vec{d}\cdot\vec{E}$$
where $\vec{d} = \sum_j e\vec{r}_j$ is the classical dipole moment of the emitter, $\vec{r}_j$ is the position of the j-the electron in this system, and e=1.6·10⁻¹⁹C. E.g., for a single electron atom, $\vec{d} = e\vec{r}$, where and $\vec{r}$ is the position of the electron relative to nucleus (and similarly, for a neutral exciton in a QD, $\vec{r}$ could be roughly interpreted as the "distance" between electron and hole). This semi-classical Hamiltonian can be converted into quantum mechanical representation (acting on atomic eigenstates |ex> and |g> and the field eigenstates |n>). First, we apply the unity operator $1 = |ex\rangle\langle ex| + |g\rangle\langle g|$ to both sides of the dipole moment:
$$e\vec{r} = 1 e\vec{r} 1$$
The resulting expression can be simplified by introducing the dipole moment matrix element: $\vec{\mu}_{ij} = \langle i|e\vec{r}|j\rangle$, and atom transition operators, defined as $\sigma_{ij} = |i\rangle\langle j|$. Transition operators for two particular cases are also referred to as the raising and lowering

operators and are denoted as $\sigma_+ = \sigma_{eg} = |ex\rangle\langle g|$ and $\sigma_- = \sigma_{ge} = |g\rangle\langle ex|$, respectively. Therefore, after this transformation we can represent the dipole moment in terms of the operators acting on the atomic states:

$$e\vec{r} = \vec{\mu}_{ee}\sigma_{ee} + \vec{\mu}_{gg}\sigma_{gg} + \vec{\mu}_{eg}\sigma_{eg} + \vec{\mu}_{ge}\sigma_{ge}$$

The final step of converting $H_{int}$ into quantum mechanical form is representing the E-field quantum mechanically, using the electric field operator:

$$\vec{E}(\vec{r}_E) = i\sqrt{\frac{\hbar\omega}{2V_{mode}\max\left\{\varepsilon(\vec{r})|\vec{E}(\vec{r})|^2\right\}}}\hat{e}\left(E(\vec{r}_E)a - E^*(\vec{r}_E)a^\dagger\right)$$

In this case, we assume that the system is single mode, $\hat{e}$ is the electric field orientation at the location $\vec{r}_E$ of the quantum emitter, $a, a^\dagger$ are annihilation and creation operators for the cavity mode, $E(\vec{r}_E)$ is the classical value of the field at the location of the emitter, and $V_{mode}$ is the cavity mode volume:

$$V_{mode} = \frac{\iiint \varepsilon(\vec{r})|\vec{E}(\vec{r})|^2 d^3\vec{r}}{\varepsilon_M|\vec{E}(\vec{r}_M)|^2}$$

In the expression for $V_{mode}$, $\vec{r}_M$ denotes the point where the electric field energy density $\varepsilon(\vec{r})|\vec{E}(\vec{r})|^2$ is maximum and $\varepsilon_M$ is the dielectric constant at this point $\varepsilon_M = \varepsilon(\vec{r}_M)$, i.e.:

$$V_{mode} = \frac{\iiint \varepsilon(\vec{r})|\vec{E}(\vec{r})|^2 d^3\vec{r}}{\max\left\{\varepsilon|\vec{E}(\vec{r})|^2\right\}}$$

By combining the expressions for dipole moment and electric field operator, we can thus express the interaction Hamiltonian as:

$$H_{int} = -i\sqrt{\frac{\hbar\omega}{2V_{mode}\max\left\{\varepsilon(\vec{r})|\vec{E}(\vec{r})|^2\right\}}}\left(\vec{\mu}_{ee}\cdot\hat{e}\sigma_{ee} + \vec{\mu}_{gg}\cdot\hat{e}\sigma_{gg} + \vec{\mu}_{eg}\cdot\hat{e}\sigma_{eg} + \vec{\mu}_{ge}\cdot\hat{e}\sigma_{ge}\right)\left(E(\vec{r}_E)a - E^*(\vec{r}_E)a^\dagger\right)$$

This expression can be simplified further by noting that $\vec{\mu}_{eg} = \vec{\mu}_{ge}^*$ and $|\vec{\mu}_{ee}| = |\vec{\mu}_{gg}| = 0$ because of the parity (for any state $|i\rangle$: $\vec{\mu}_{ii} = \langle i|e\vec{r}|i\rangle = e\langle i|\vec{r}|i\rangle = 0$, since the position operator is an odd function). After multiplying all of the remaining terms:

$$H_{int} = -i\sqrt{\frac{\hbar\omega}{2V_{mode}\max\left\{\varepsilon(\vec{r})|\vec{E}(\vec{r})|^2\right\}}}\begin{pmatrix}\vec{\mu}_{eg}\cdot\hat{e}E(\vec{r}_E)\sigma_+ a + \vec{\mu}_{eg}^*\cdot\hat{e}E(\vec{r}_E)\sigma_- a - \\ \vec{\mu}_{eg}\cdot\hat{e}E^*(\vec{r}_E)\sigma_+ a^\dagger - \vec{\mu}_{eg}^*\cdot\hat{e}E^*(\vec{r}_E)\sigma_- a^\dagger\end{pmatrix}$$

The terms proportional to $\sigma_- a$ and $\sigma_+ a^\dagger$ do not conserve energy and they can be neglected. Therefore, we are left with the following interaction Hamiltonian:

$$H_{int} = -i\sqrt{\frac{\hbar\omega}{2V_{mode}\max\left\{\varepsilon(\vec{r})|\vec{E}(\vec{r})|^2\right\}}}\left(\vec{\mu}_{eg}\cdot\hat{e}E(\vec{r}_E)\sigma_+ a - \vec{\mu}_{eg}^*\cdot\hat{e}E^*(\vec{r}_E)\sigma_- a^\dagger\right)$$

Let's introduce *the emitter-field coupling parameter* $g(\vec{r})$:

$$g(\vec{r}_E) = \frac{1}{\hbar}\sqrt{\frac{\hbar\omega}{2V_{\text{mode}}\max\{\varepsilon(\vec{r})|\vec{E}(\vec{r})|^2\}}}\vec{\mu}_{eg}\cdot\hat{e}E(\vec{r}_E)$$

Then we can express the interaction Hamiltonian in the following form:
$$H_{\text{int}} = i\hbar\left(g^*(\vec{r}_E)a^\dagger\sigma_- - g(\vec{r}_E)\sigma_+ a\right)$$

The coupling parameter $g(\vec{r}_E)$ is the product of the Rabi frequency $g_0$, a position dependent part $\psi(\vec{r}_E)$, and a polarization dependent part $\cos(\xi)$:
$$g(\vec{r}_E) = g_0\psi(\vec{r}_E)\cos(\xi),$$
where
$$g_0 = \frac{\mu_{eg}}{\hbar}\sqrt{\frac{\hbar\omega}{2\varepsilon_M V_{\text{mode}}}}, \quad \psi(\vec{r}_E) = \frac{E(\vec{r}_E)}{|E(\vec{r}_M)|}, \quad \text{and} \quad \cos(\xi) = \frac{\vec{\mu}_{eg}\cdot\hat{e}}{\mu_{eg}}.$$

and $\mu_{eg} = |\vec{\mu}_{eg}|$.

Therefore, the coupling strength $|g(\vec{r}_E)|$ reaches its maximum value of $|g_0|$ when the emitter is located at the point $\vec{r}_M$ where the electric field energy density is maximum, and when its dipole moment is aligned with the electric field, i.e., when $\psi(\vec{r}_E) = 1$ and $\cos(\xi) = 1$. To maximize $|g_0|$, we should pick the cavity mode with as small mode volume $V_{\text{mode}}$ as possible. This describes the motivation behind using photonic crystal cavities for cavity QED experiments.

To summarize, the total Hamiltonian of the coupled emitter-cavity field system is described with:
$$H = H_A + H_F + H_{\text{int}},$$
where $H_A = \frac{\hbar\nu}{2}\sigma_z$, $H_F = \hbar\omega\left(a^\dagger a + \frac{1}{2}\right)$ and $H_{\text{int}} = i\hbar\left(g^*(\vec{r}_E)a^\dagger\sigma_- - g(\vec{r}_E)\sigma_+ a\right)$. The Hamiltonian H is called the *Jaynes-Cummings Hamiltonian*, and describes a lossless emitter-field system (Kimble, Cavity Quantum Electrodynamics 1994).

### 3.2. Strong coupling regime: Rabi oscillation and anharmonic dressed states ladder

Without the interaction turned on ($H_{\text{int}}=0$), the total Hamiltonian of the emitter-field system is:
$$H_0 = \frac{\hbar\nu}{2}\sigma_z + \hbar\omega\left(a^\dagger a + \frac{1}{2}\right)$$

The eigenstates of this Hamiltonian are |ex,n> and |g,n+1>, i.e., the product of atomic and field eigenstates. The corresponding eigenenergies are sums of the atomic and field eigenenergies:

$$E_1 = \frac{\hbar\nu}{2} + \hbar\omega\left(n + \frac{1}{2}\right) = \frac{\hbar\delta}{2} + \hbar\omega(n+1)$$

$$E_2 = -\frac{\hbar\nu}{2} + \hbar\omega\left(n + \frac{3}{2}\right) = -\frac{\hbar\delta}{2} + \hbar\omega(n+1).$$

When detuning is not present ($\delta = \nu - \omega = 0$), it follows that the energy levels are degenerate and equal to $E_0 = \hbar\omega(n+1)$.

When the interaction Hamiltonian is turned on, these states are coupled, and the evolution of the state of the system can be found by solving the time-dependent Schrödinger equation:

$$H|\psi(t)\rangle = i\hbar\frac{d|\psi(t)\rangle}{dt}$$

$$|\psi(t)\rangle = \left(C_{e,n}(t)|ex,n\rangle + C_{g,n+1}(t)|g,n+1\rangle\right)e^{-i(n+1)\omega t}$$

where $\delta = \nu - \omega$ is the detuning between the cavity field frequency and the atomic transition frequency. When detuning is not present ($\delta = \nu - \omega = 0$), and assuming that the system is initially in the state |ex,n> at t=0, the solution to the coupled system of equations is:

$$C_{g,n+1}(t) \propto \sin\left(|g(\vec{r}_E)|\sqrt{n+1}\,t\right)$$

$$C_{e,n}(t) \propto \cos\left(|g(\vec{r}_E)|\sqrt{n+1}\,t\right)$$

The probabilities of the system being in the state |g,n+1> and |ex,n> are, respectively:
$|C_{g,n+1}(t)|^2 = \frac{1 - \cos\left(2|g(\vec{r}_E)|\sqrt{n+1}\,t\right)}{2}$ and $|C_{e,n}(t)|^2 = \frac{1 + \cos\left(2|g(\vec{r}_E)|\sqrt{n+1}\,t\right)}{2}$. Therefore, in the strong coupling regime, the time-evolution of the system can be described with oscillation at the frequency $2|g(\vec{r}_E)|\sqrt{n+1}$ between the states |ex,n> and |g,n+1>. This process is called the *Rabi oscillation*. For the case when the system is initially in the state |ex,0>, i.e., no photons are present in the cavity and the atom is in the excited state, the frequency of the Rabi oscillation is equal to $2|g(\vec{r}_E)|$.

The coupled emitter-field system (with interaction turned on) has new eigenstates. These new eigenstates can be found by solving for the eigenvectors and eigenvalues of the Hamiltonian including the interaction. The representation of the total Hamiltonian of the coupled emitter-cavity field system in the basis {|ex,n>,|g,n+1>} is:

$$H = \begin{bmatrix} \langle ex,n|H|ex,n\rangle & \langle ex,n|H|g,n+1\rangle \\ \langle g,n+1|H|ex,n\rangle & \langle g,n+1|H|g,n+1\rangle \end{bmatrix}$$

$$H = \begin{bmatrix} \frac{\hbar\delta}{2} + \hbar\omega(n+1) & -i\hbar g\sqrt{n+1} \\ i\hbar g^*\sqrt{n+1} & -\frac{\hbar\delta}{2} + \hbar\omega(n+1) \end{bmatrix}$$

and eigenenergies (eigenvalues of this Hamiltonian) are:

$$E_{\pm} = \hbar\omega(n+1) \pm \sqrt{\left(\frac{\hbar\delta}{2}\right)^2 + \hbar^2|g|^2(n+1)} \quad \ldots (3.2.1)$$

When detuning between the emitter and the field is not present ($\delta=0$), it follows that:
$$E_{\pm} = \hbar\omega(n+1) \pm \hbar|g|\sqrt{(n+1)}$$
and the corresponding eigenstates are:

$E_+$: $|n+1,+\rangle = \dfrac{|ex,n\rangle + |g,n+1\rangle}{\sqrt{2}}$ (even mode)

$E_-$: $|n+1,-\rangle = \dfrac{|ex,n\rangle - |g,n+1\rangle}{\sqrt{2}}$ (odd mode)

States $|n+1,+\rangle$ and $|n+1,-\rangle$ are termed *dressed states* (*normal modes*), and states $|ex,n\rangle$, $|g,n+1\rangle$ are termed *bare states*. Dressed states are eigenstates of the total Hamiltonian, including interaction (Fig. 7 and Fig. 8). Bare states are eigenstates of the Hamiltonian without the interaction turned on. If the system is prepared in the bare state and the interaction Hamiltonian is turned on, the Rabi oscillation between the bare states occurs, as we have shown previously.

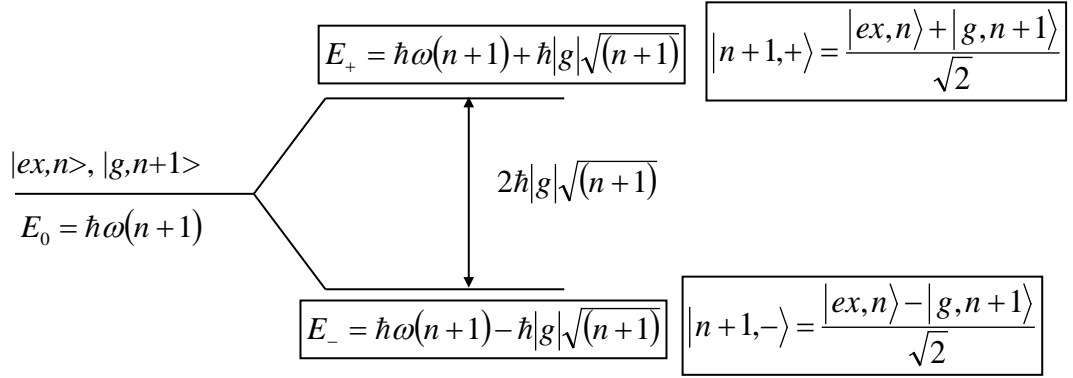

**Figure 7.** Splitting of the eigenstates of uncoupled emitter-field system into dressed states once the emitter-field interaction is turned on. n corresponds to the number of photons in the field mode, and $|ex\rangle$, $|g\rangle$ are the excited and ground states of the emitter, as shown in Fig. 6.

Therefore, if the atom-field interaction is turned on and detuning between the cavity field frequency and the atom transition frequency is not present, the energy level $E_0$ splits into two energy levels $E_+$ and $E_-$, corresponding to the even and odd modes, respectively. This splitting is usually used as the indication that the atom-cavity system has reached the strong coupling regime. Such splitting happens for any excitation level (integer n>0, as shown in Fig. 7), and therefore the new structure of eigenstates is represented by a ladder shown in Fig. 8.

The dressed states ladder in Figure 8 is plotted only for zero detuning between the atom and the cavity. When detuning $\delta$ is not zero, the eigenenergies as a function of detuning would be represented by plotting the expression for $E_\pm$ with $\delta$ given above (Eq. 3.2.1), leading to anticrossing curves shown in Figure 9 (Fabrice P. Laussy 2012).

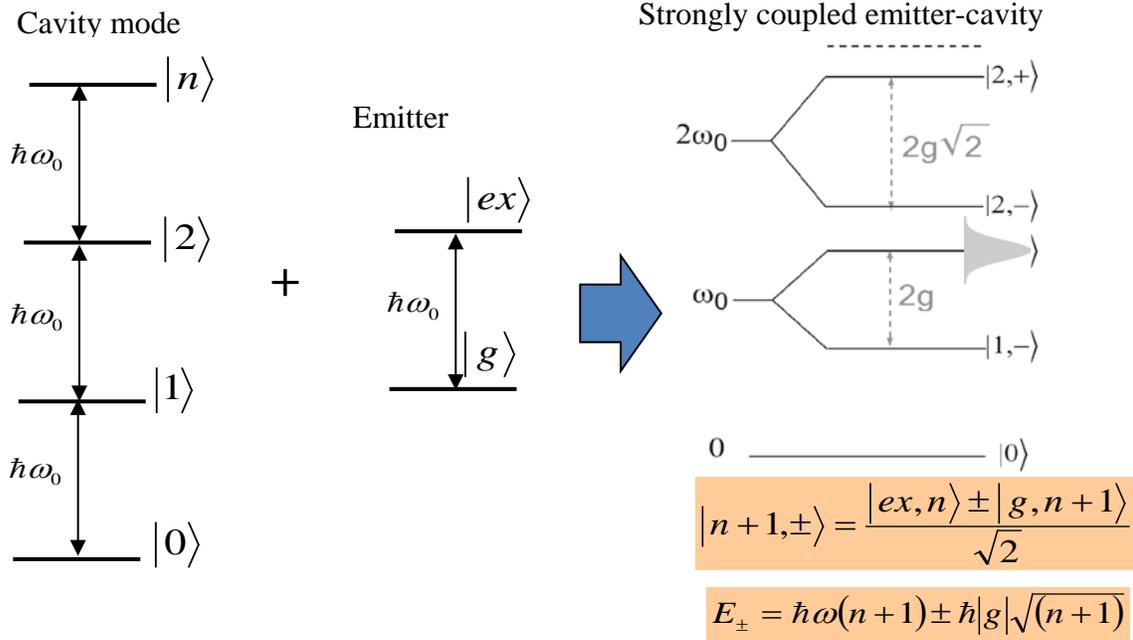

**Figure 8.** Cavity mode has energy eigenstates described by a harmonic ladder (left) while a single 2-level emitter has energy eigenstates |ex> and |g> (middle). Once they are strongly coupled, new energy eigenstates are formed called dressed states (right), and the energy eigenvalues are positioned on an anharmonic ladder called the dressed states ladder.

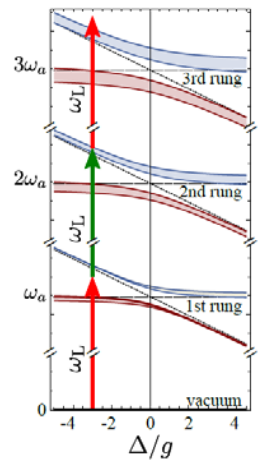

**Figure 9.** Dressed states ladder for a detuned emitter-cavity system (courtesy of Prof. Fabrice Laussy, Universidad Autónoma de Madrid, reproduced from (Fabrice P. Laussy 2012)).

As can be seen in Figures 8 and 9, the ladder of dressed states is anharmonic. In other words, the splitting between dressed state energy levels is not constant. This can be employed to demonstrate the effects such as *photon blockade*, where the coupling of the second photon with the frequency $\omega_0 \pm g$ into the cavity QED system is prohibited, as only one photon with that frequency can occupy one of the first two dressed states (see Figure 10). This effect was first demonstrated in atomic cavity QED system (K. M. Birnbaum 2005) and then in quantum dot cavity QED system. (Andrei Faraon 2008) (Andreas Reinhard 2012)

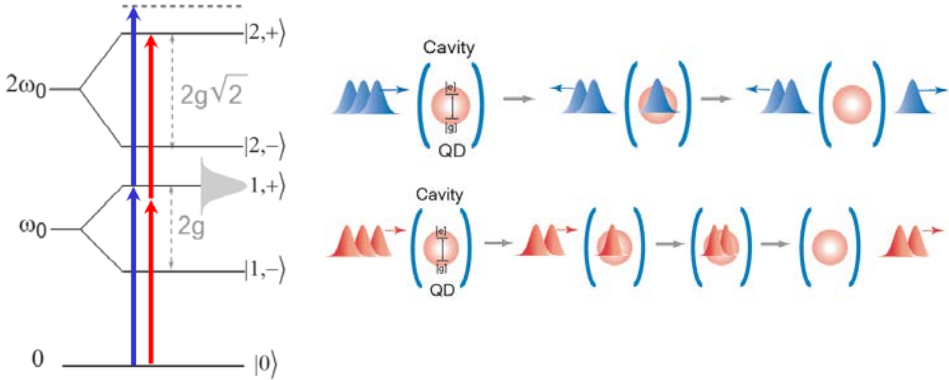

**Figure 10.** The excitation one of the eigenstates in the first manifold of the dressed states ladder can be done with frequency $\omega_0+g$ (in a tuned system), shown with the blue arrow. One can place only one excitation (one photon) into this first manifold, as it is a result of mixing between 1 photon and 1 atom. However, the distance from this eigenstate to any eigenstates in the second manifold is different from $\omega_0+g$ (because of the ladder anharmonicity). Therefore, only one photon from the excitation can be coupled to the system, and once it is coupled, all the other will be repelled (as in the top-right cartoon). This effect is referred to as the *photon blockade*. Similarly, if the excitation has frequency given by the red arrow, then an eigenstate in the second manifold is reached (where 2 photons can be placed). In this case, only pairs of photons can be coupled to the system, as shown in the cartoon on bottom right. This effect is referred to as the *photon induced tunneling*.

Similarly, higher order manifolds in the ladder can be probed, leading to filtering of n-photon states via the effect called *photon induced tunneling*. In an example shown in Figure 10, photon pairs are filtered by exciting the second manifold. This effect was demonstrated in solid state (quantum dot-cavity system) (Andrei Faraon 2008) and in atomic physics (I. Schuster 2008).

Usually, eigenstates are probed in transmission of a probe laser through the system. For example, if we had an empty cavity (no strongly coupled emitter inside it), it would feature the transmission spectrum as in Fig 11 (left). However, if an emitter is loaded into the cavity and strongly coupled to the cavity field, the cavity transmission spectrum changes. If the excitation is weak and only first manifold of the ladder can be accessed, the transmission spectrum features two peaks, with the splitting equal to $2|g|$ (as in Fig. 11 - right). These two peaks correspond to the two eigenstates in the first manifold. With stronger excitation, additional peaks start popping in the transmission spectrum (positioned in between the two initial peaks) corresponding to higher order manifolds (I. Schuster 2008).

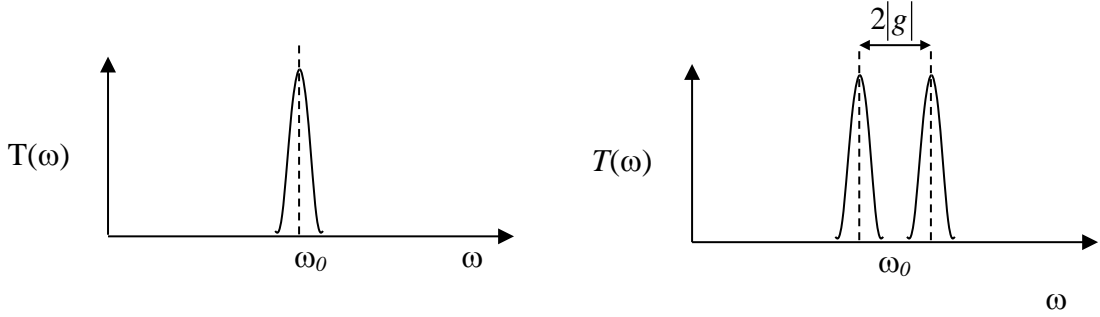

**Figure 11.** Rabi splitting. **Left**: Transmission spectrum of an empty cavity mode. **Right**: Transmission spectrum of a strongly coupled emitter-cavity system for weak excitation, when only eigenstates in the first manifold can be accessed.

### 3.3. Strongly coupled cavity QED system with losses

Here we will follow a very simple semi-classical model to include the effect of losses in a cavity QED system. Recall that the eigenfrequencies (eigenvalues of the Jaynes-Cummings Hamiltonian) for the first manifold of the system without losses are:

$$E_{\pm} = \hbar\omega \pm \sqrt{\left(\frac{\hbar\delta}{2}\right)^2 + \hbar^2|g|^2}$$

here $\delta = \nu - \omega$ (atom-cavity detuning). This expression was derived by setting the ground state for field at $\hbar\omega/2\,\nu$ and for atom at $-\hbar\nu/2$, as described in Section 3.1. If we set both to zero, the energy eigenvalues are:

$$\frac{E_{\pm}}{\hbar} = \omega_{\pm} = \omega - \frac{\omega}{2} + \frac{\nu}{2} \pm \sqrt{\left(\frac{\delta}{2}\right)^2 + |g|^2} = \frac{(\omega+\nu)}{2} \pm \sqrt{\left(\frac{\delta}{2}\right)^2 + |g|^2}$$

Let us now assume that the system has losses: the emitter can decay into other channels (e.g. spontaneous emission into other modes, or nonradiative decay) with rate $\gamma$ and the cavity is imperfect, so that the cavity mode can lose photons with cavity field decay rate $\kappa$. In other words, $\kappa$ and $\gamma$ are half-width half-maxima of the cavity and emitter spectra, respectively.

To account for losses, we will replace emitter and cavity mode frequencies with complex frequencies:

$$\nu \to \nu - i\gamma$$
$$\omega \to \omega - i\kappa$$

Now let us plug these complex frequencies back into the expression for eigenfrequencies of the strongly coupled system:

$$\omega_{\pm} = \frac{(\omega+\nu)}{2} - i\frac{(\kappa+\gamma)}{2} \pm \sqrt{\left(\frac{\delta - i(\kappa-\gamma)}{2}\right)^2 + |g|^2}$$
……(3.3.1)

These are eigenfrequencies of the first manifold in the cavity QED system with losses (J. P. Reithmaier 2004). The real part of $\omega_{\pm}$ gives frequencies of eigenmodes, and the imaginary part describes their damping.

Let us assume that we have a good emitter, with linewidth $\gamma$ much narrower than cavity linewidth $\kappa$. We also assume that the emitter-cavity coupling strength g is larger than any losses in the system (g> $\kappa$/2; $\kappa$>>$\gamma$), and that emitter-cavity mode detuning $\delta$=0. To reach this regime, one has to increase g relative to $\kappa$, which can be achieved by reducing cavity mode volume $V_{mode}$ (thereby increasing g) and by increasing cavity Q factor (thereby decreasing $\kappa$). In this regime, the expression under the square root is positive and we have new eigenfrequencies:

$$\omega_{\pm} = \omega \pm \sqrt{|g|^2 - \frac{\kappa^2}{4}} - i\frac{\kappa}{2}$$

Therefore, we expect to see two distinct lines in the spectrum with frequencies $\omega_{\pm} = \omega \pm \sqrt{|g|^2 - \frac{\kappa^2}{4}}$ that have full widths half maxima of $\kappa$. The system is strongly coupled, exhibits Rabi splitting, as shown in Figure 11, but the splitting between the two peaks is not exactly 2g as in the lossless system, and the peaks are broadened by cavity decay. In the case when g>> $\kappa$/2, the splitting approaches 2g, as in the lossless system:

$$\omega_{\pm} = \omega \pm |g| - i\frac{\kappa}{2}$$

The two peaks correspond to new eigenstates of the system, which are the entangled states of the emitter and the cavity field, as shown in Fig. 8.

### 3.4. Weakly coupled cavity QED system – Purcell regime

Let us now consider a different regime of emitter-cavity field interaction in which the cavity field decay rate dominates: ($\kappa$/2 >>g>>$\gamma$) and detuning $\delta$=0. In the expression for $\omega_{\pm}$ (Eq. 3.3.1), the expression under the square root is now negative. Therefore, both eigenstates are at the same frequency $\omega$ (as real parts of $\omega_{\pm}$ are the same), but their dampings are different. The damping (i.e., imaginary part of $\omega_{\pm}$) is:

$$\frac{\kappa}{2} \mp \sqrt{\frac{\kappa^2}{4} - |g|^2} = \frac{\kappa}{2}\left(1 \mp \sqrt{1 - \frac{4|g|^2}{\kappa^2}}\right) \approx \frac{\kappa}{2}\left(1 \mp \left(1 - \frac{1}{2}\frac{4|g|^2}{\kappa^2}\right)\right) \approx \begin{cases}|g|^2/\kappa \\ \kappa\end{cases}$$

Therefore, in this regime we expect two eigenstates with the same frequency ($\omega$) but different damping; their energy would decay with rates $2|g|^2/\kappa$ and $2\kappa$, respectively. This is the so-called *Purcell regime* – a special situation in the weak coupling regime of cavity QED: there would be no Rabi splitting in the spectrum, and emitter and cavity would just cross, but the decay rate (spontaneous emission rate) of the emitter would be modified by the presence of the cavity. In this regime, one eigenstate (cavity mode) has the energy decay rate of $2\kappa$. The other eigenstate (at the same frequency $\omega$) is the emitter, with modified radiative decay rate equal to $2|g|^2/\kappa$. This modified spontaneous emission rate is very different from the spontaneous emission rate of that same emitter without the cavity, in bulk (uniform material with refractive index n), where it is equal to $\Gamma_n = n\Gamma_0$, and $\Gamma_0 = \frac{\mu_{eg}^2 \nu^3}{3\pi\varepsilon_0 \hbar c^3}$ is the Einstein A coefficient (Zubairy 1997).

The ratio between the spontaneous emission rate of an emitter in a cavity and the same emitter without the cavity is the Purcell factor:

$$F = \frac{2|g|^2/\kappa}{n\Gamma_0}$$

Clearly, the Purcell factor reaches its maximum value when the emitter is on resonance with the cavity and spatially aligned with the cavity field maximum (when $g=g_0$). By plugging in the expressions for $g_0$ (see section 3.1.), $\kappa$, and $\Gamma_0$ into the expression above, we obtain that the maximum Purcell factor is:

$$F = \frac{3}{4\pi^2}\left(\frac{\lambda}{n}\right)^3 \frac{Q}{V_{\mathrm{mode}}}$$

If the emitter is not spatially aligned to the cavity mode maximum, $g = g_0 \psi(\vec{r})\cos(\xi)$, as shown in Section 3.1., where $\psi$ and cos describe its spatial offset from the E-field energy density maximum and cosine describes misalignment between its dipole moment and E-field polarization at that location. The reduction in *g* also degrades the Purcell enhancement. Finally, the previous analysis was done for an emitter than is tuned to the cavity resonance, i.e., $\delta=\nu-\omega=0$. If there is a spectral detuning between the two, the Purcell factor F is also degraded, in proportion with the density of optical states

$$D_c(\nu) = \frac{1}{\pi} \frac{\omega/2Q}{(\nu-\omega)^2 + (\omega/2Q)^2}.$$

Therefore, the spontaneous emission is not an intrinsic property of a quantum emitter (e.g., atom or exciton), but is rather a property of an emitter coupled to its electromagnetic environment (Purcell 1946). The spontaneous emission rate is directly proportional to the density of electromagnetic states, and can be modified with respect to its value in free space by placing the emitter in a cavity. Inside a cavity, this rate can be significantly enhanced in proportion to $Q/V_{\mathrm{mode}}$. Similarly, by decreasing the density of optical states (e.g., inside a photonic bandgap), one can reduce the emission rate. (Yablonovitch 1987)

## 3.5. Spontaneous emission coupling factor and relation to lasing threshold

In the previous section, we showed that the spontaneous emission rate of an emitter (e.g., an atom, or a quantum dot exciton) can be increased by increasing the $Q/V_{mode}$ ratio of a cavity. We will now show that the increase in this ratio also leads to the reduction of the lasing threshold.

Consider an emitter inside a cavity and with an enhanced spontaneous emission rate $\Gamma_c = F\Gamma_0$ (as a result of coupling to a cavity mode), where F is the Purcell factor and $\Gamma_0$ is the spontaneous emission rate of that same emitter without a cavity (e.g., in free space, or bulk). Let us denote by $\Gamma_{other}$ the spontaneous emission rate of this emitter into all modes other than the cavity mode. We can express $\Gamma_{other} = f\Gamma_0$, where f<1 in photonic band gap, but otherwise $f \approx 1$. The total spontaneous emission rate is $\Gamma_{total} = \Gamma_{other} + \Gamma_c$. The fraction of the light emitted by an emitter that is coupled into one particular cavity mode is known as the spontaneous emission coupling factor β:

$$\beta = \frac{\Gamma_c}{\Gamma_{total}} = \frac{\Gamma_c}{\Gamma_c + \Gamma_{other}} = \frac{F}{F+f}$$

Therefore, if the spontaneous emission rate of an emitter is strongly enhanced by its interaction with a cavity mode (F>>1), or of the emission to other modes is significantly suppressed (f<<1), the spontaneous emission coupling factor $\beta \approx 1$. This means that nearly all spontaneous emission goes into the lasing mode. The fraction of spontaneous emission going into non-lasing modes is one of the fundamental losses in a laser, and by decreasing it one can lower the laser threshold. Therefore by increasing β we can decrease the lasing threshold.

Let us now prove this conclusion from the simple laser rate equations. We denote the number of photons in the cavity mode by *p*, and the total number of emitters in the excited state by *N*. An excited emitter can decay into the ground state by three mechanisms: a) it emits a photon to modes other than the cavity mode with rate $\Gamma_{other}$ (spontaneous emission); b) it emits a photon into the cavity mode with rate $\Gamma_{cav}(p+1)$ (stimulated emission); and c) it decays without emission of photon with rate $1/\tau_{nr}$ (non-radiative decay). It is also assumed that the emitters are pumped back to the excited state with rate $R_{pump}$. Therefore:

$$\frac{dN}{dt} = R_{pump} - N\Gamma_{other} - N\Gamma_{cavity}(p+1) - \frac{N}{\tau_{nr}} \quad \ldots(3.5.1)$$

The last term can be neglected, under the assumption that the nonradiative decay rate is small. Similarly, the rate equation for the number of photons in the cavity mode is:

$$\frac{dp}{dt} = -2\kappa p + N\Gamma_{cavity}(p+1) \quad \ldots(3.5.2)$$

where $\kappa = \frac{\omega}{2Q}$ is the cavity field decay rate. Clearly, the number of photons increases as a result of the emission rate into the cavity mode, and decreases as a result of the cavity loss.

The equations (3.5.1) and (3.5.2) constitute the rate equations for this system. If we solve them in a steady state (d/dt=0), express N in terms of p from the 2$^{nd}$ equation, and plug that back into the 1$^{st}$ equation, we obtain:

$$R_{pump} = \frac{2\kappa p}{\Gamma_{cavity}(p+1)}\left(\Gamma_{other} + \Gamma_{cavity}(p+1)\right) \ldots(3.5.3)$$

On the other hand, we know that:

$$\beta = \frac{\Gamma_{cavity}}{\Gamma_{cavity} + \Gamma_{other}} \Rightarrow \frac{1}{\beta} - 1 = \frac{\Gamma_{other}}{\Gamma_{cavity}} \quad \ldots(3.5.4)$$

By plugging (3.5.4) into (3.5.3):

$$R_{pump} = \frac{2\kappa p}{p+1}\left(\frac{1}{\beta} + p\right) \quad \ldots(3.5.5)$$

The lasing threshold is defined as the condition when the mean photon number in a cavity is equal to 1, i.e., $p=1$ (Imamoglu 1999). Therefore, above threshold $p>1$, i.e., the stimulated emission occurs (as there is more than one photon to stimulate emission). From Eq. (3.5.5), it follows that the pump rate at the lasing threshold is:

$$R_{pump} = \frac{\kappa}{\beta}(\beta+1) = \frac{\omega}{2Q\beta}(\beta+1) = \frac{\omega}{2Q}\left(1+\frac{1}{\beta}\right)$$

For β<<1, as is the case in conventional lasers $R_{pump,th} \approx \frac{\omega}{2Q\beta}$.

Therefore, the lasing threshold pump rate decreases with an increase in the spontaneous emission coupling factor β and the cavity Q-factor. In other words, by employing smaller volume and higher-Q cavities, one can achieve lasing at lower pumping powers.

## 4. Quantum optics and cavity QED experiments with quantum dots in photonic crystal cavities

### 4.1. Rabi splitting and anticrossing

As an illustration of the strong coupling regime with a lossy cavity and a good emitter ($g>\kappa/2$ and $\kappa>>\gamma$), let us consider a single InAs/GaAs quantum dot (QD) exciton in a GaAs photonic crystal cavity, as shown in Fig. 3. (A. F. Dirk Englund 2007) (A. M. Dirk Englund 2010). Figure 12 shows the result of the cross-polarized reflectivity measurement on a strongly coupled QD cavity system which operates in this regime. Cross polarized reflectivity measurement is described in details in our earlier work (A. F. Dirk Englund 2007), but here we only note that such measurement leads to the result identical to the transmission measurement through the system. Figure 12-left shows the anticrossing curve (for another QD-cavity system) where QD is tuned across the cavity resonance by changing sample temperature. A clear Rabi splitting is visible in both cases. From the experimental results we extract that a typical range of parameters in our strongly coupled QD-cavity system is $g/2\pi$=10-25GHz, $\kappa/2\pi$=8-16GHz, and $\gamma/2\pi$~1GHz. The use of very small mode volume photonic crystal cavities enables operation in the ultrafast g regime, and reaching of the strong coupling regime ($g>\kappa/2$) even with the moderate values of the cavity Q factor of around Q~$10^4$.

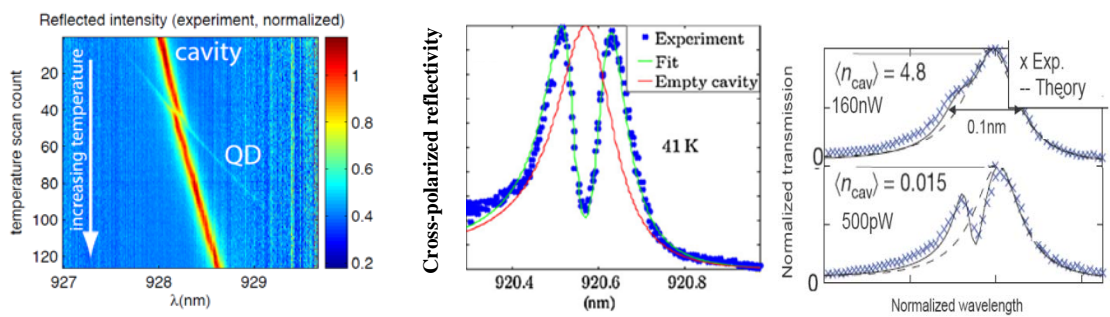

**Figure 12. Left:** Mapping the anticrossing curve (the first manifold) of the strongly coupled QD-photonic crystal cavity system by changing the sample temperature. **Middle**: Rabi splitting of the photonic crystal cavity resonance after a QD exciton is strongly coupled to it. **Right:** At higher power of the probe laser, the strongly coupled system saturates and Rabi splitting disappears.

It is worth noting that at higher probe laser powers (roughly exceeding 1 photon per cavity lifetime), the dip in the split spectrum (and Rabi splitting) disappears (see Fig. 12-right). This is the result of the system saturation. Classically, the probe laser sends enough photons to keep the emitter constantly excited, so for all other incoming photons the system behaves as an empty cavity. Quantum mechanically, at higher probe laser powers one can excite higher manifolds of the dressed state ladder (Fig. 8) which appear in the spectrum as peaks at frequencies between the two peaks of the first manifold. Therefore, these higher order peaks fill up the dip in the Rabi split spectrum and the system is saturated.

### 4.2. Controlled amplitude and phase shifts; switching with a strongly coupled system

It is possible to employ the described saturation effects to perform controlled amplitude and phase switching using the strongly coupled system. Let us shine two beams on the system (instead of one); they should be at the same frequency, or slightly detuned from each other, but both still within the cavity resonance. In this case, both beams can jointly saturate the system. We will refer to one of them as signal, and the other as control. If only a signal beam is present, its power is insufficient to saturate the system. Since its frequency coincides with the dip of the strongly coupled system, the beam is just reflected back (see Fig. 13 – left). On the other hand, if the signal beam is accompanied by a control beam (Fig. 13 – right), they jointly saturate the system, the Rabi splitting disappears, and both beams are transmitted to the other side. The structure thus behaves as an AND logic gate for two beams.

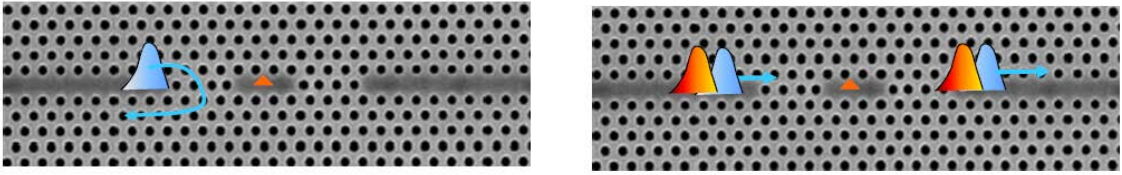

**Figure 13.** Switching with a strongly coupled cavity QED system. Signal beam only (left) has insufficient power to saturate the system and is reflected back (as its frequency is within the dip of the Rabi split system), but if it accompanied with the control beam (right) they saturate the system and are transmitted together to the other side.

Fig. 14 shows the results of such experiment with a single QD strongly coupled to a PC cavity, and with two continuous wave (CW) beams that are slightly detuned from each other (Ilya Fushman 2008). The signal beam is very weak and cannot saturate the system on its own. When control beam photon number in the cavity approaches one ($n_c \sim 1$), the system is saturated and amplitude and phase of the transmitted signal beam look like those for an empty cavity. Therefore, with only a few nW of power in the control beam ($n_c \sim 1$, when saturation happens), both amplitude and phase of the transmitted signal beam are completely changed. For complete saturation, phase shifts of up to $\pi/4$ are achievable.

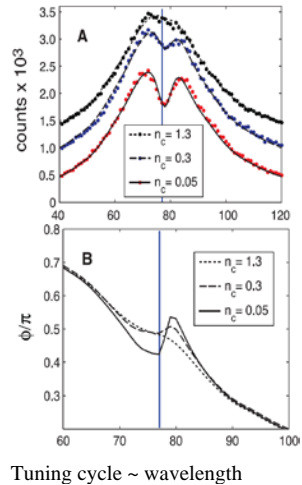

Tuning cycle ~ wavelength

**Figure 14.** Controlled amplitude (A) and phase (B) shifts with a single QD strongly coupled to a photonic crystal cavity. The plots map the amplitude and phase of the probe beam at the output as a function of the control beam photon number $n_c$. In saturation amplitude and phase curves look like the corresponding curves of an empty cavity. Both control and probe beams are continuous wave and slightly detuned from each other.

For practical applications, it is desirable to perform such switching with optical pulses. The results of such experiments with a single QD strongly coupled to a photonic crystal cavity are shown in Fig. 15 (A. M. Dirk Englund 2012). Signal and control beams consist of 40ps pulses and they are resonant with each other. Individually, they can't saturate the system, but jointly they can. We map transmission through the system as a function of delay between the signal and control pulses. When the delay is greater than the pulse duration, there is no saturation, and pulses bounce back (as in Fig 13 - left). However, when the delay is comparable or smaller than the pulse duration, the system is saturated and pulses propagate together to the other side (like in Fig. 13-right). As can be seen in Fig. 15d, the transmission rises when the pulse delay becomes comparable or smaller than the pulse duration. Therefore, with a strongly coupled QD-cavity system, we can perform all optical switching at 40GHz speed and with optical powers on the order of few nW. Similar switching experiments were also demonstrated by (Ranojoy Bose 2012) and (Thomas Volz 2012).

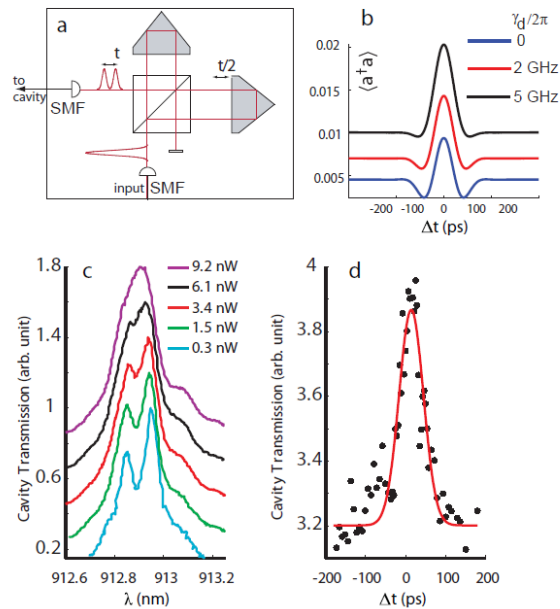

**Figure 15.** Interaction of two weak laser pulses through the QD/cavity system. **(a)** Time-delay setup for producing pulses at a separation of $\Delta t$. **(b)** Simulated interaction of two laser pulses, represented by the instantaneous intracavity photon number $\langle a^\dagger a \rangle$ as a function of the time delay $\Delta t$ between the two 40 ps long Gaussian pulses. Curves are calculated for a set of different rates of pure QD dephasing, $\gamma_d$, which causes a reduction of the transmission dips before and after the peak. Pure dephasing also causes a blurring of the spectral normal mode splitting, which in turn raises the transmission for increasing $\gamma_d$. **(c)** Pump-power dependence of the cavity transmission for coincident pulses repeating at 80 MHz. **(d)** Signal observed when the cavity-QD system is probed with two 40 ps pulses as a function of their delay. When the two pulses have a temporal overlap inside the cavity, the QD saturates and the overall cavity reflection increases. The power in the single of the two pulses corresponds roughly to the 3.4nW trace in (c). Best agreement is found with the theoretical plot for a pure dephasing rate $\gamma_d/2\pi \sim 5$ GHz.

### 4.3. Generation of nonclassical light with the strongly coupled quantum dot-cavity system: photon blockade and tunneling

The effects of photon blockade and photon induced tunneling resulting from the anharmonicity of the dressed states ladder were previously explained in Section 3.2. As shown in Figure 10, such effects can be used to generate nonclassical light (Fock states) by transmission of the laser beam at the right frequency through a strongly coupled cavity QED system. Photon blockade was first demonstrated in an atomic cavity QED system (K. M. Birnbaum 2005), and higher order manifolds were also probed in an atomic cavity QED system (I. Schuster 2008). Here we present the results of photon blockade and photon induced tunneling in a strongly coupled quantum dot cavity QED system (Andrei Faraon 2008) .

To show that in laser transmission we are generating a certain non-classical state of light, we are measuring a 2$^{nd}$ order photon correlation function g$^{(2)}$($\tau$) (Zubairy 1997) using a Hanburry Brown and Twiss type setup shown in Fig. 16:

$$g^{(2)}(\tau) = \frac{\langle a^\dagger a^\dagger(\tau) a(\tau) a \rangle}{\langle a^\dagger a \rangle^2}$$

where $n(t) = a^\dagger(t)a(t)$ is the photon number operator, $a = a(0)$, and $a^\dagger = a^\dagger(0)$. Let us assume that the transmitted state through the system can be expresses a superposition of the Fock states: $|\psi\rangle = \sum_n c_n |n\rangle$, where the probability of the n-th Fock state is $P_n = |c_n|^2$. Then (by employing $a^\dagger|n\rangle = \sqrt{n+1}|n+1\rangle$ and $a|n\rangle = \sqrt{n}|n-1\rangle$) it follows that:

$$g^{(2)}(0) = \frac{\langle \psi | a^\dagger a^\dagger a a | \psi \rangle}{\langle \psi | a^\dagger a | \psi \rangle^2} = \frac{\sum_n n(n-1)P_n}{\left(\sum_n n P_n\right)^2}$$

Using this expression, let us evaluate g$^{(2)}$(0) for several cases of interest:

1) *perfect single-photon pulse train*: $P_n = \begin{cases} 1, & n = 1 \\ 0, & n \neq 1 \end{cases}$

   It is simple to show that g$^{(2)}$(0)=0

2) *sparse single photon pulse train* (mostly vacuum with sporadic single photon pulses):
   $P_n = \begin{cases} \varepsilon, & n = 1 \\ 1-\varepsilon, & n = 0 \\ 0, & n \neq 0,1 \end{cases}$, where $\varepsilon \ll 1$

   By plugging this into the expression for g$^{(2)}$(0), we can show that g$^{(2)}$(0)=0 still holds.

3) *perfect two-photon pulse train*: $P_n = \begin{cases} 1, & n = 2 \\ 0, & n \neq 2 \end{cases}$

   In this case, g$^{(2)}$(0)=1/2

4) *perfect N-photon pulse train*: $P_n = \begin{cases} 1, & n = N \\ 0, & n \neq N \end{cases}$

In this case, $g^{(2)}(0)=(N-1)/N$

5) *sparse train of two-photon pulses*: $P_n = \begin{cases} \varepsilon, & n = 2 \\ 1-\varepsilon, & n = 0 \\ 0, & n \neq 0,2 \end{cases}$

In this case, $g^{(2)}(0)=1/(2\varepsilon)$

Since $\varepsilon \ll 1$ for a sparse train, it follows that $g^{(2)}(0)>1$ in this case, i.e., we expect to observe bunching.

6) *sparse train of three-photon pulses*: $P_n = \begin{cases} \varepsilon, & n = 3 \\ 1-\varepsilon, & n = 0 \\ 0, & n \neq 0,3 \end{cases}$

In this case, $g^{(2)}(0)=2/(3\varepsilon)$

And again, we expect bunching ($g^{(2)}(0)>1$), as $\varepsilon \ll 1$ for a sparse train.

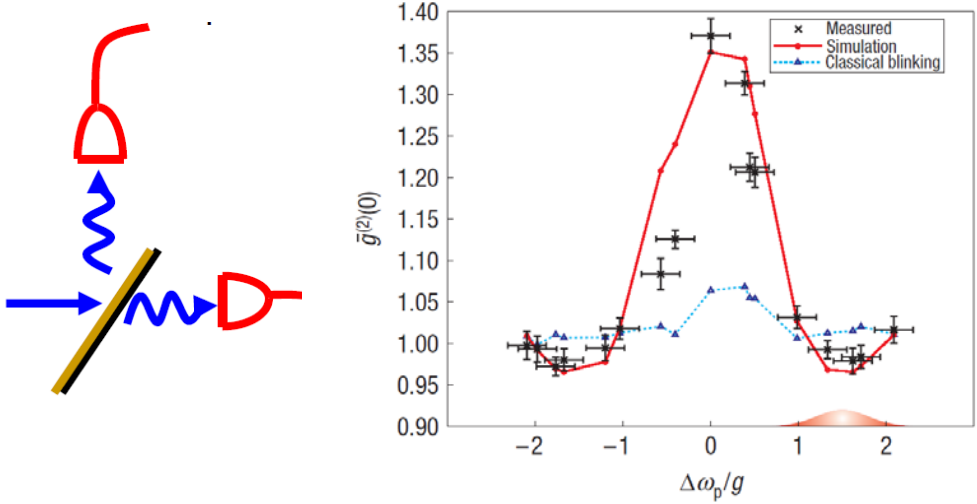

**Figure 16. Left:** Hanburry Brown and Twiss setup for $g^{(2)}(t)$ measurements consists of a beamsplitter and two single photon counters. Right: Measurement of $g^{(2)}(0)$ for transmission of a laser beam trough a strongly coupled QD-photonic crystal cavity system. The horizontal axis corresponds to the tuning of the input laser frequency relative to the empty cavity frequency. The transmission through an empty cavity should exhibit a Poissonian statistics (like a laser itself: $g^{(2)}(0)=1$). However, in this case we observe the regimes of antibunching ($g^{(2)}(0)<1$) corresponding to sub-Poissonian statistics (imperfect single photon stream in the regime of photon blockade), and superpoissionian statistics (bunching $g^{(2)}(0)>1$ ) in the regime of photon induced tunneling.

Similarly, if the output state is a combination of these it can be shown that $g^{(2)}(0)$ varies between antibunching (sub-poissonian regime, with $g^{(2)}(0)<1$) and bunching (super-poissonian regime, with $g^{(2)}(0)>1$). For a coherent state (at the output of the laser) which features poissonian statistics of photons, it can be shown that $g^{(2)}(0)=1$ (Zubairy 1997).

The measurement of $g^{(2)}(0)$ for a laser beam transmitted through a strongly coupled QD-cavity system as a function of the laser detuning from the empty cavity frequency is shown in Fig. 16 (Andrei Faraon 2008). Clearly, $g^{(2)}(0)$ goes between antibunching (in the photon blockade regime, where sub-poissonian light is generated, i.e., the output is a superposition of Fock states with dominant single photon component) and bunching (in the photon induced tunneling regime where the output is mostly vacuum with sparse higher order Fock states.

**4.4. Strong coupling of single quantum dot to a photonic molecule**

As an extension of the single quantum dot - photonic crystal cavity QED experiments shown in Section 4.1., we now show strong coupling experiments between a single QD and a photonic crystal molecule (i.e., the system of two coupled cavities) (Arka Majumdar 2012).

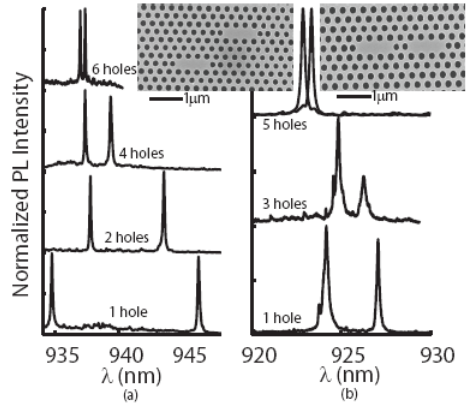

**Figure 17.** Spectra of photonic crystal molecules as a function of the separation of coupled cavities. As expected, the increase in separation between the cavities decreases their coupling strength and reduces the spectral separation between the coupled modes.

A photonic molecule consists of two coupled photonic crystal cavities in this case, as shown In Fig. 17. The spectrum of a such a system is expected to exhibit two coupled modes corresponding to bonding and anti-bonding states, and whose separation increases as the coupling between the cavities increases (which in turn can be done by decreasing their spatial separation – see Fig. 17).

Let us consider a system of coupled cavities as in Fig. 17-left with a single QD inside of it. If the system consists of three strongly coupled oscillators (2 modes and 1 QD), it is expected that the modes would form supermodes and the QD would anticross both of them. Indeed, by tuning a QD across both supermodes (by combination of temperature and nitrogen condensation tuning), we observe its anticrossing with both supermode peaks, as shown in Fig. 18. Such a system has potential applications in nonclassical light generation (Savona 2010).

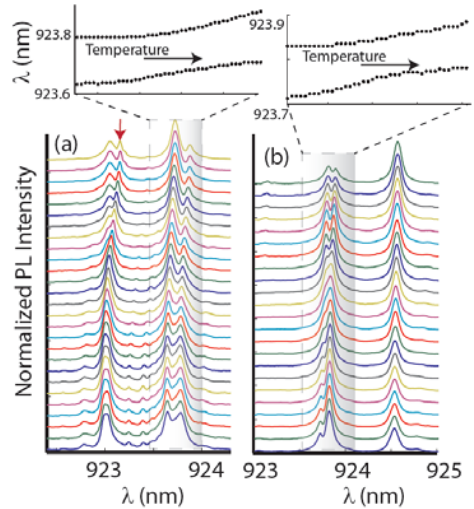

**Figure 18.** Strong coupling of a single InAs/GaAs quantum dot to a photonic crystal molecules, like the one shown in Fig. 17-left. Anticrossing of the QD with both supermodes is observed, as it is tuned across them. Therefore, the system consists of three strongly coupled oscillators.

### 4.5. Weak coupling of a single quantum dot to a cavity: Purcell regime and spontaneous emission rate suppression

In the Purcell regime of cavity QED described in Section 3.4., we expect to observe enhancement of the spontaneous emission rate of a quantum emitter as a result of its coupling to a cavity resonance. This enhancement scales as $Q/V_{mode}$ of a cavity, and is maximum for an emitter that is spectrally and spatially aligned with the cavity. If the emitter is spectrally detuned, the Purcell enhancement follows a lorentzian lineshape given by the density of optical states in a cavity. The experiments shown in Fig. 19 indeed demonstrate such a result for a single InAs/GaAs QD in a DBR micropost cavity (basically 1D photonic crystal cavity) (J. Vuckovic 2003)**.**

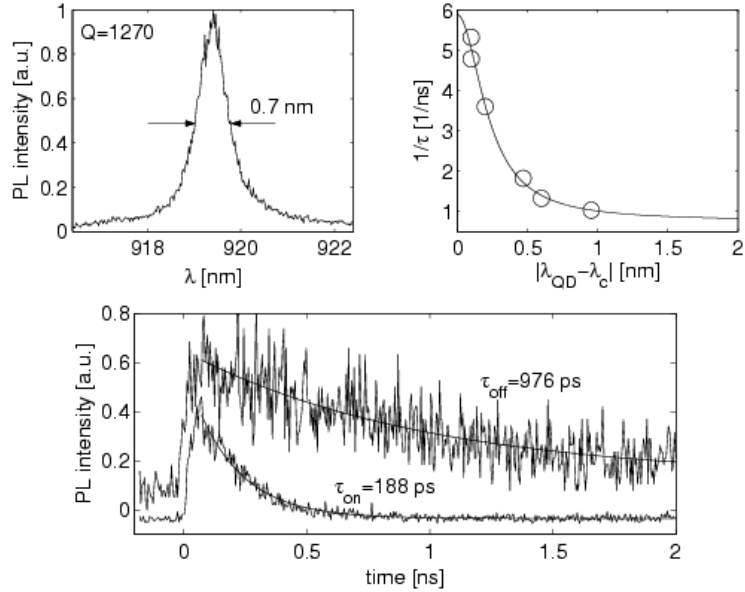

**Figure 19.** Weak coupling of a single QD to a DBR micropost cavity. **Top left:** cavity Q factor. **Top right:** QD spontaneous emission rate as it is tuned across the cavity resonance follows the lorentzian lineshape described by the cavity spectrum. Bottom: Time resolved lifetime measurements for this QD, showing the maximum Purcell effect (shortest lifetime), and longest lifetime.

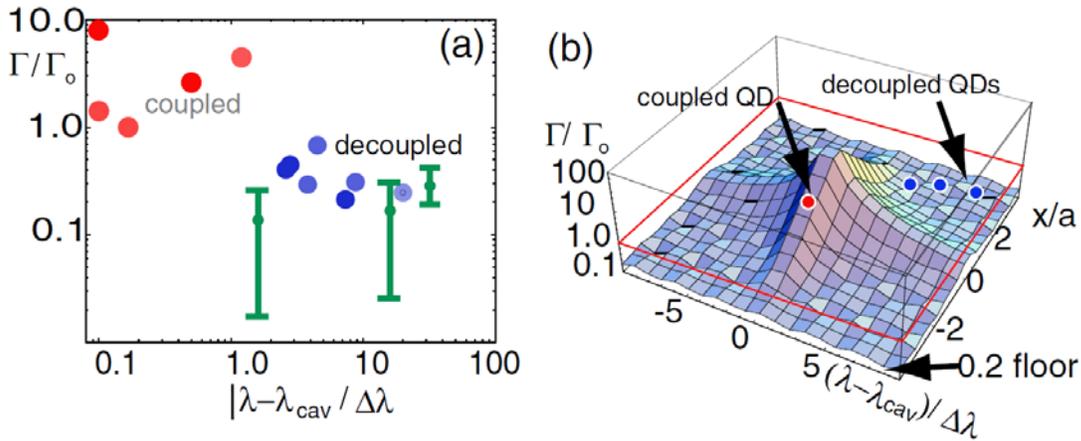

**Figure 20.** (a) Spontaneous emission rate modification of individual QDs as a function of their detuning from 2D photonic crystal cavity resonance. Tuned QD exhibit Purcell enhancement or not much enhancement (depending on whether they are spatially aligned with the cavity), while spectrally detuned QDs always exhibit suppression of the spontaneous emission, as a result of the reduction of the photon density of states in photonic band gap Green bars show simulation results. (b) The predicted spontaneous emission rate modification in the photonic crystal cavity as a function of normalized spatial and spectral misalignment from the cavity ($a$ is the lattice periodicity). This plot assumes $Q = 1000$ and polarization matching between the emitter dipole and cavity field.

The results shown in Fig. 19 confirm significant Purcell enhancement for a QD resonant with the cavity mode, and show that the Purcell factor as a function of QD detuning from the cavity indeed follows a lorentizan lineshape given by the cavity spectrum.

Similarly, as described in Section 3.4., if the density of photon states $D_c(v)$ is zero or significantly suppressed, the spontaneous emission rate should also be suppressed (as $\Gamma/\Gamma_0 \propto D_c(v)$). This is for example, possible in photonic crystals, as a result of photonic band gap, as shown in Fig. 20 (D. F. Dirk Englund 2005).

### 4.6. Ultralow threshold photonic crystal laser

In the Section 3.5 we learned that the Purcell enhancement also leads to the enhancement in the spontaneous emission coupling factor β and the reduction in lasing threshold. This was employed in optically pumped photonic crystal lasers with InAs quantum dots as an active medium to demonstrate β~0.85 and lasing threshold power of 124nW  (S. Strauf 2006).

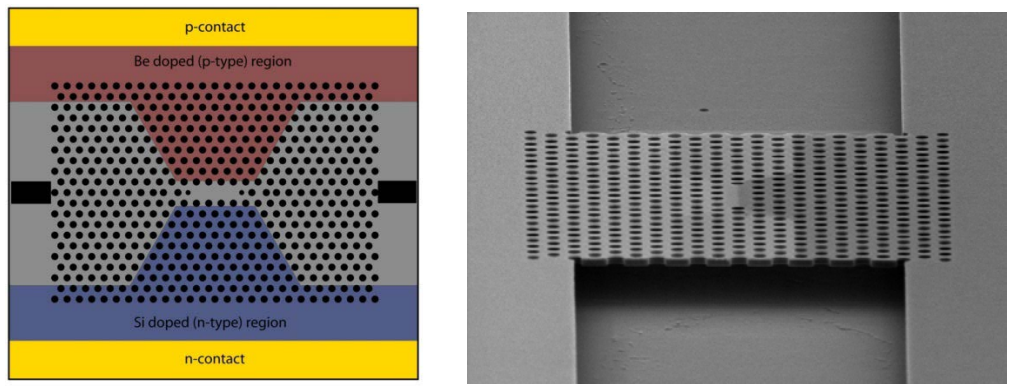

**Figure 21. Left**: lateral p-i-n junction embedded in a photonic crystal cavity for electrical injection into the cavity region. **Right**: such a cavity fabricated in GaAs with InAS quantum dots as an active medium.

However, for practical applications it is necessary to inject lasers electrically. This was recently achieved in the same GaAs photonic crystal cavity lasers with an ensemble of QDs as an active medium, using the configuration of lateral pin junction shown in Fig. 21 (Bryan Ellis 2011).  The demonstration of an ultralow threshold electrically injected laser in this configuration are shown in Fig. 22, with lasing threshold current as low as 180nA (with 1V bias voltage).

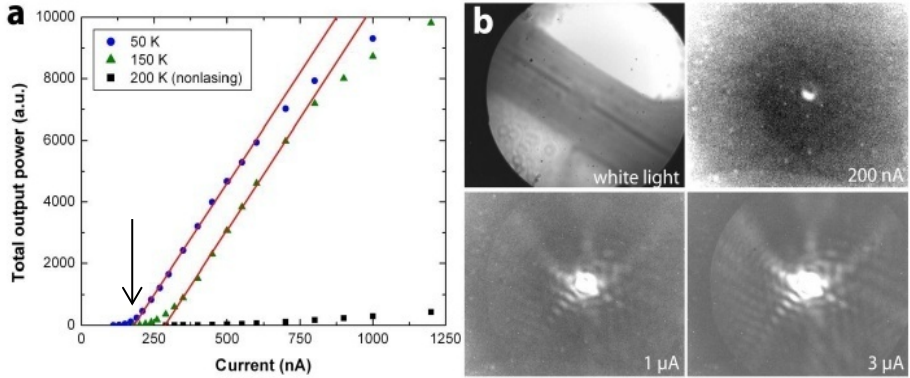

**Figure 22. (a)** Laser optical output power – injection current characteristics at different temperatures. At 50K, the lasing threshold is as low as 180nW. **(b)** Optical image of the laser and lasing patterns at different injection currents.

## 5. Summary

This chapter provided the introduction into quantum optics and solid state cavity QED with quantum dots in photonic crystal cavities. Although the focus was on the physics of the quantum dot – photonic crystal cavity system, there are many applications that would greatly benefit from this platform. In addition to the already mentioned all optical switches at a few photon level and ultra-low threshold lasers, combining such blocks into circuits can be used to construct all optical computers, optical interconnects in computers, or quantum repeaters for long distance quantum communication (Kimble, The quantum internet 2008), (Jeremy O'Brien 2009). Finally, although we focused on a single InAs/GaAs quantum dot as a quantum emitter inside a planar GaAs photonic crystal cavity (D. F. Dirk Englund 2005) (T. Yoshie 2004) (A. Badolato 2005) (A Kress 2005), the physics described here applies to other types of quantum emitters, including those in solid state – such as NV centers in diamond coupled to cavities (B. S. Dirk Englund 2010) (Faraon 2013) (A Faraon 2011), quantum dots in other types of structures (Benson 2011) (A. Laucht 2012) (Lodahl 2012) (J. P. Reithmaier 2004) (D. Press 2007) (Painter 2007) (Kanna Aoki 2008), neutral atoms in microcavities (P. Maunz 2005) (J. D. Thompson 2013) (K. M. Birnbaum 2005) and even circuit QED systems (A. Wallraff 2004)

## 6. Acknowledgment

I would like to thank all of the present and former members of my research group who have worked on the described cavity QED and quantum optics projects with quantum dots in photonic crystals: Prof. Dirk Englund (MIT), Prof. Andrei Faraon (Caltech), Dr. Ilya Fushman (Dropbox), Prof. Edo Waks (University of Maryland and JQI), Prof. Vanessa Sih (University of Michigan), Prof. Arka Majumdar (University of Washington, Seattle), Dr. Erik Kim (HGST), Prof. Michal Bajcsy (IQC, Waterloo), Dr. Bryan Ellis (Soraa), Dr. Gary Shambat (Adamant Technologies), Dr. Per Kaer (DTU),

Dr. Konstantinos Lagoudakis, Dr. Tomas Sarmiento, Dr. Kai Mueller, Mr. Armand Rundquist, Mr. Kevin Fisher, and Mr. Jan Petykiewicz. I would also like to thank Prof. Yoshi Yamamoto (Stanford) and my former co-workers from his research group, Dr. Charlie Santori and Dr. David Fattal (HP Labs) with whom I performed Purcell effect measurements on a single QD in a micropost cavity, as well as Prof. Pierre Petroff (emeritus, UCSB), Prof. Jim Harris (Stanford) and Prof. Seth Bank (UT Austin) for growth of quantum dots used in some of these experiments. In addition, I would like to thank ARO (Dr. TR Govindan), AFOSR (Dr. Tatjana Curcic and Dr. Gernot Pomrenke), ONR (Dr. Chagaan Baatar), DARPA (Dr. Jag Shah), and NSF (Dr. Wendy Fuller-Mora and Dr. Dan Finotello) for funding these projects. Finally, I would like to thank the organizers of this summer school, Prof. Claude Fabre, Prof. Nicolas Treps, and Prof. Vahid Sandoghdar, as well as many school participants for many stimulating discussions at Les Houches.

Bryan Ellis, Marie Mayer, Gary Shambat, Tomas Sarmiento, Eugene Haller, James S. Harris, and Jelena Vuckovic. "Ultralow-threshold electrically pumped quantum-dot photonic-crystal nanocavity laser." *Nature Photonics*, 2011: 297-300.

D. Press, S. Goetzinger, S. Reitzenstein, C. Hofmann, A. Loeffler, M. Kamp, A. Forchel, and Y. Yamamoto. " Photon Antibunching from a Single Quantum Dot-Microcavity System in the Strong Coupling Regime." *Physical Review Letters*, 2007: 117402.

Dirk Englund, Andrei Faraon, Ilya Fushman, Nick Stoltz, Pierre Petroff, and Jelena Vuckovic. "Controlling Cavity Reflectivity With a Single Quantum Dot." *Nature*, 2007: 857-861.

Dirk Englund, Arka Majumdar, Andrei Faraon, Mitsuru Toishi, Nick Stoltz, Pierre Petroff, and Jelena Vuckovic. "Resonant excitation of a quantum dot strongly coupled to a photonic crystal nanocavity." *Physical Review Letters* , 2010: 073904.

Dirk Englund, Arka Majumdar, Michal Bajcsy, Andrei Faraon, Pierre Petroff, and Jelena Vuckovic. "Ultrafast Photon-Photon Interaction in a Strongly Coupled Quantum Dot-Cavity System." *Physical Review Letters*, 2012: 093604 .

Dirk Englund, Brendan Shields, Kelley Rivoire, Fariba Hatami, Jelena Vuckovic, Hongkun Park, and Mikhail D. Lukin. "Deterministic coupling of a single nitrogen vacancy center to a photonic crystal cavity." *Nano Letters*, 2010: 3922-3926.

Dirk Englund, David Fattal, Edo Waks, Glenn Solomon, Bingyang Zhang, Toshihiro Nakaoka, Yasuhiko Arakawa, Yoshihisa Yamamoto, and Jelena Vuckovic. "Controlling the Spontaneous Emission Rate of Single Quantum Dots in a 2D Photonic Crystal." *Physical Review Letters*, 2005: 013904.

Fabrice P. Laussy, Elena del Valle, Michael Schrapp, Arne Laucht and Jonathan J. Finley. "Climbing the Jaynes–Cummings ladder by photon counting." *Journal of Nanophotonics*, 2012: 061803.

Faraon, Marko Loncar and Andrei. "Quantum photonic networks in diamond." *MRS Bulletin*, 2013: 144-148.

Haroche, Serge. "Nobel Lecture: Controlling photons in a box and exploring the quantum to classical boundary." *Reviews of Modern Physics*, 2013: 1083–1102.

I. Schuster, A. Kubanek, A. Fuhrmanek, T. Puppe, P. W. H. Pinkse, K. Murr and G. Rempe. "Nonlinear spectroscopy of photons bound to one atom." *Nature Physics*, 2008: 382-385 .

Ilya Fushman, Dirk Englund, Andrei Faraon, Nick Stoltz, Pierre Petroff, and Jelena Vuckovic. "Controlled Phase Shifts with a Single Quantum Dot." *Science*, 2008: 769-772.

Imamoglu, Yoshihisa Yamamoto and Atac. *Mesoscopic quantum optics*. New York: John Wiley and Sons, 1999.

J. D. Thompson, T. G. Tiecke, N. P. de Leon, J. Feist, A. V. Akimov, M. Gullans, A. S. Zibrov, V. Vuletić, and M. D. Lukin. "Coupling a Single Trapped Atom to a Nanoscale Optical Cavity." *Science*, 2013: 1202-1205.

J. P. Reithmaier, G. Seogonk, A. Löffler, C. Hofmann, S. Kuhn, S. Reitzenstein, L. V. Keldysh, V. D. Kulakovskii, T. L. Reinecke and A. Forchel. "Strong coupling in a single quantum dot–semiconductor microcavity system." *Nature*, 2004: 197-200.
35